\documentclass[twocolumn,showpacs,preprintnumbers,amsmath,amssymb]{revtex4}
\usepackage{dcolumn,graphicx}
\usepackage{bm}
\usepackage[latin1]{inputenc}

\DeclareMathOperator{\Conc}{Conc} \DeclareMathOperator{\tr}{tr}
\begin{document}
\preprint{}
\title{A criterion for entanglement in two two-level systems}
\author{E. Ferraro, A. Napoli, A. Messina}
\affiliation{Dipartimento di Scienze Fisiche ed Astronomiche,
Università di Palermo, via Archirafi 36, 90123 Palermo, Italy}
\begin{abstract}
We prove a necessary and sufficient condition for the occurrence of
entanglement in two two-level systems, simple enough to be of
experimental interest. Our results are illustrated in the context of
a spin star system analyzing the exact entanglement evolution of the
central couple of spins.
\end{abstract}


\maketitle It is today experimentally possible to realize many
entangled states in different contexts like CQED
\cite{Haroche}-\cite{Saif}, metal-superconductor junctions
\cite{Bayandin}, \cite{Kim} and other solid state systems
\cite{Bose1}, \cite{Bose2}. A very debated and open issue concerns
the individuation of efficient methods to detect and quantify
entanglement. In principle, tomography techniques allow the
reconstruction of the state of the system and to establish whether
the system is entangled with the help of some separability criteria.
To determine in laboratory the density operator matrix elements
requires however considerable efforts, making thus desirable to
probe and to quantify the occurrence of entanglement without using a
tomographic approach. To this end Bell inequalities and witness
operators \cite{Horodecki}-\cite{Wu} could be used even if their
exploitation is indeed still an hard task. It is indeed an exciting
challenge to find criteria for detecting entanglement involving the
measurement of few observables of clear physical meaning. Methods
based on violation of local uncertainty relations \cite{Hofmann} or
inequalities for variances of observables \cite{Guhne} have been
quite recently proposed. In this paper we present a criterion for
entanglement well suited for bipartite systems in which each
component is a two-level system. Our results do provide an incisive
tool to infer the existence of non classical correlations from few
and simple measurements. Our criterion is successfully applied to a
spin star system.

Consider a bipartite system composed by two two-level systems
described by the Pauli operators
$\vec{\sigma}_i\equiv(\sigma_{\pm}^{(i)}, \sigma_z^{(i)})\quad i=1,
2$ respectively. Suppose that its density operator, in the
factorized basis $\{\vert\!\!
\uparrow\uparrow\rangle,|\!\!\uparrow\downarrow\rangle,
|\!\!\downarrow\uparrow\rangle,\\|\!\!\downarrow\downarrow\rangle\}$
of the eigenstates of $\sigma_z^{(1)}\sigma_z^{(2)}$, has the
following structure
\begin{equation}\label{struttura}
\rho=\left(%
\begin{array}{cccc}
  a & 0 & 0 & 0 \\
  0 & b & c & 0 \\
  0 & c^\ast & d & 0 \\
  0 & 0 & 0 & e \\
\end{array}%
\right).
\end{equation}
The quite simple form of $\rho$, given by eq.(\ref{struttura})
naturally arises in many physical scenarios not necessarily
involving spin $\frac{1}{2}$ systems
\cite{Eberly,Yu,Pratt,Bayandin}. Since the concurrence function
associated to eq.(\ref{struttura}) is
\begin{equation}\label{conc}
\Conc=\max\bigl[0,2\bigl(|c|-\sqrt{a\,e}\bigr)\bigr]
\end{equation}
the existence of entanglement in our system requires that the
geometric mean of the two populations $a$ and $e$ becomes less than
the coherence amplitude $|c|$. The two probabilities $a$ and $e$ of
finding the system in the state $|\!\!\uparrow\uparrow\rangle$ and
$|\!\!\downarrow\downarrow\rangle$ respectively, determine the mean
value of $S_z\equiv\frac{1}{2}(\sigma_z^{(1)}+\sigma_z^{(2)})$ and
$S_z^2$ as follows
\begin{equation}\label{s1} \langle
S_z\rangle\equiv\tr\{\rho S_z\}=a-e,
\end{equation}
\begin{equation}\label{s2}
\langle\bigl(S_z\bigr)^2\rangle\equiv\tr\{\rho (S_z)^2\}=a+e
\end{equation}
which in turn imply that
\begin{equation}\label{fluc}
\bigl(\Delta S_z\bigr)^2\equiv\langle
\bigl(S_z\bigr)^2\rangle-\bigl(\langle
S_z\rangle\bigr)^2=(a+e)-(a-e)^2.
\end{equation}
In accordance with the Landau's condition \cite{Landau},
\begin{equation}\label{landau}
|c|\leq\sqrt{b\,d},
\end{equation}
we may thus state that the presence of entanglement in the two
two-level systems described by eq.(\ref{struttura}), necessarily
requires
\begin{equation}\label{condizione}
\sqrt{a\,e}<|c|\leq \sqrt{b\,d}.
\end{equation}
Exploiting such an inequality in eq.(\ref{fluc}) yields the
following upper limit to the variance of $S_z$:
\begin{equation}\label{cn}
\bigl(\Delta S_z\bigr)^2<4b\,d-(b+d)[(b+d)-1].
\end{equation}
It is of relevance the fact that when the density matrix
(\ref{struttura}) assumes the special form
\begin{equation}\label{struttura2}
\rho(t)=\left(%
\begin{array}{cccc}
  a & 0 & 0 & 0 \\
  0 & b & b & 0 \\
  0 & b & b & 0 \\
  0 & 0 & 0 & e \\
\end{array}%
\right)
\end{equation}
the necessary condition (\ref{cn}) simply reduces to
\begin{equation}\label{cns1}
\bigl(\Delta S_z\bigr)^2<2b
\end{equation}
or, equivalently,
\begin{equation}\label{cns}
\bigl(\langle S_z\rangle\bigr)^2>1-4b.
\end{equation}
Since when condition (\ref{cns}) is fulfilled then the concurrence
is different from zero, we may claim that eq.(\ref{cns}) is a
necessary and sufficient condition in order to have entanglement in
two two-level systems described by $\rho$ as given by
eq.(\ref{struttura2}). Stated another way the occurrence of
entanglement in the system may be checked simply comparing the
square of the mean value of $S_z$ with the population $b$. It is
however important to stress that the result we have obtained allows
us to immediately deduce that if $b>\frac{1}{4}$, then we can be
sure that the two two-level systems are entangled. On the contrary
when $b\leq\frac{1}{4}$ we need also to know the mean value of
$S_z$. In this case indeed the bipartite system is entangled if, and
only if,
\begin{equation}\label{disu}
-|2b-1|\leq\langle S_z\rangle<-\sqrt{1-4b}
\end{equation}
or, alternatively,
\begin{equation}
\sqrt{1-4b}<\langle S_z\rangle\leq|2b-1|.
\end{equation}
In other words when $b\leq\frac{1}{4}$, the presence of entanglement
in the system is compatible only with $S_z$ mean values 'squashed'
toward the extremes of its interval of variability $[-1,1]$. In turn
it implies that under the condition $0 < b\leq\frac{1}{4}$, in order
to have entanglement, the probability of finding the system in the
state $|\!\!\uparrow\uparrow\rangle$ must be appreciably different
from the probability to find the system in the state
$|\!\!\downarrow\downarrow\rangle$.

\begin{figure}[h]
\begin{center}
\includegraphics[scale=0.7]{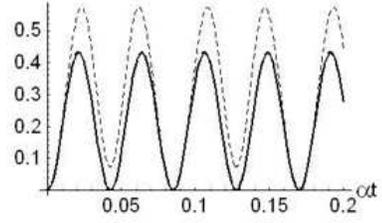}
\end{center}
\caption{ $(\Delta S_z)^2$ (dashed line) and $2b(t)$ (bold line)
against $\alpha t$ in correspondence to $k=2$ and
$N=100$}\label{Immagine7a}
\end{figure}

Our criterion may be successfully exploited in order to analyze the
entanglement evolution in a spin system describing a physical
scenario of interest in many physical contexts. Consider indeed a
system constituted by two uncoupled spins $\frac{1}{2}$, denoted by
$A$ and $B$ and hereafter called central system, each one
interacting with $M-2$ mutually uncoupled spins $\frac{1}{2}$. In
particular we suppose that the two central spins interact with each
of the $M-2$ spins in the same way. This system, known as spin star
like system \cite{Bose,Petruccione}, can be described by adopting
the following hamiltonian model
\begin{equation}\label{Hamiltonian}
H=H_0+H_I
\end{equation}
with
\begin{equation}
H_0=\omega(S_z+J_z),
\end{equation}
\begin{equation}
H_I=\alpha(S_+J_-+S_-J_+),
\end{equation}
where $S_z$ and
$S_\pm\equiv\frac{1}{2}(\sigma_{\pm}^{(1)}+\sigma_{\pm}^{(2)})$ are
spin operators acting in the Hilbert space of the central system and
$J_z$ and $J_\pm$ are the collettive spin operators describing the
other $M-2$ spins. This hamiltonian model can be successfully used
to describe for example electronic spins in semiconductor quantum
dot coupled by hyperfine interaction with nuclear spins, or
electronic spins bound to phosphorus atoms in a matrix of silica or
germanio in presence of defects \cite{Schliemann}. The hamiltonian
model (\ref{Hamiltonian}) possesses permutational symmetries
successfully exploitable to exactly solve the relative
time-dependent Schr\"{o}dinger equation \cite{Palumbo},
\cite{Napoli}. Suppose that at $t=0$ the central system is in a
common eigenstate of $S^2=(\vec{S}_1+\vec{S}_2)^2$ and $S_z$ denoted
by $|S,M_S\rangle$. At the same time the remaining $M-2$ spins are
supposed in the state $|J,M_J,\nu\rangle$, eigenstate of the
collettive angular momentum operators $J^2$ and $J_z$. The index
$\nu$, depending on $J$, allows us to distinguish between different
states of the coupled angular momentum basis characterized by the
same $J$ and $M_J$. Hamiltonian (\ref{Hamiltonian}) is invariant by
permutation of the two central spins as well as of an arbitrary
couple of spins among the $M-2$ of the second block. Moreover
$[S^2,H]=[J^2,H]=[S_z+J_z,H]=[J^2_{int},H]=0$, $\vec{J}_{int}$ being
an intermediate angular momentum resulting from the coupling of
selected at will individual angular momentum of the $M-2$ spins. At
a generic time instant $t$ we can write
\begin{equation}\label{evolution}
|\psi(t)\rangle=e^{-iH_0t}e^{-iH_It}|S,M_S\rangle|J,M_J,\nu\rangle,
\end{equation}
being $[H_0,H_I]=0$. The case $S=0$ and consequently $M_S=0$, is a
trivial one corresponding to an eigenstate of the Hamiltonian
whatever the value of $J$ and $M_J$ are. When instead $S=1$ we have
\begin{widetext}\begin{equation}\label{hi}
\begin{split}
H_I^{2n}|1,M_S\rangle|J,M_J,\nu\rangle=&\left[\alpha^{2n}(\sqrt{2})^{2n}p_{M_S}(p_{M_S}^2+r_{M_S}^2)^{n+\frac{(-1)^{M_S}-1}{2}}\right]\Bigl(p_{M_S}|1,M_S\rangle|J,M_J,\nu\rangle+\\
&+r_{M_S}|1,-M_S\rangle|J,M_J+2M_S,\nu\rangle\Bigr)
\end{split}
\end{equation}
and \begin{equation}\begin{split}
&H_I^{2n+1}|1,M_S\rangle|J,M_J,\nu\rangle=\\
&=\begin{cases}\alpha^{2n+1}(\sqrt{2})^{2n+1}p_{M_S}(p_{M_S}^2+r_{M_S}^2)^n|1,0\rangle|J,M_J+M_S,\nu\rangle & \text{if $M_S\neq0$,}\\
\alpha^{2n+1}(\sqrt{2})^{2n+1}\left[p_{-1}(p_{-1}^2+p_1^2)^n|1,1\rangle|J,M_J-1,\nu\rangle+p_1(p_{-1}^2+p_1^2)^n|1,-1\rangle|J,M_J+1,\nu\rangle\right]&
\text{if $M_S=0$}\end{cases}
\end{split}\end{equation}\end{widetext}
with
\begin{equation}\begin{split}
&p_s=\sqrt{J(J+1)-M_J(M_J+s)},\quad s=\pm1 \\
&p_0=1,
\end{split}\end{equation}
\begin{equation}\label{hf}\begin{split}
&r_s=\sqrt{J(J+1)-(M_J+s)(M_J+2s)},\quad s=\pm1 \\
&r_0=0.
\end{split}\end{equation}
After straightforward calculations we thus get
\begin{equation}\label{evolution3}
\begin{split}
&|\psi(t)\rangle=e^{-i\omega(M_S+M_J)t}\left\{A_{J\,M_J}^{M_S}(t)|1,M_S\rangle|J,M_J,\nu\rangle+\right.\\
&+B_{J\,M_J}^{M_S}(t)|1,-M_S\rangle|J,M_J+2M_S,\nu\rangle\\
&-i\left(\frac{1-(-1)^{M_S}}{2}\right)C_{J\,M_J}^{M_S}(t)|1,0\rangle|J,M_J+M_S,\nu\rangle\\
&-i\delta_{M_S\,0}\left[D_{J\,M_J}^{M_S}(t)|1,1\rangle|J,M_J-1,\nu\rangle+\right.\\
&\left.\left.+E_{J\,M_J}^{M_S}(t)|1,-1\rangle|J,M_J+1,\nu\rangle\right]\right\}.
\end{split}\end{equation}
with
\begin{equation}\begin{split}\label{A}
A_{J\,M_J}^{M_S}(t)=&\left[\frac{p_{M_S}^2}{p_{M_S}^2+r_{M_S}^2}\cos\Bigl(\sqrt{2(p_{M_S}^2+r_{M_S}^2)}\alpha
t\Bigr)+\right.\\
&+\left.\frac{r_{M_S}^2}{p_{M_S}^2+r_{M_S}^2}\right],
\end{split}\end{equation}
\begin{equation}\label{B}
B_{J\,M_J}^{M_S}(t)=\frac{p_{M_S}r_{M_S}}{p_{M_S}^2+r_{M_S}^2}\Bigl(\cos\bigl(\sqrt{2(p_{M_S}^2+r_{M_S}^2)}\alpha
t\bigr)-1\Bigr),
\end{equation}
\begin{equation}\label{C}
C_{J\,M_J}^{M_S}(t)=\frac{p_{M_S}}{\sqrt{p_{M_S}^2+r_{M_S}^2}}\sin\Bigl(\sqrt{2(p_{M_S}^2+r_{M_S}^2)}\alpha
t\Bigr),
\end{equation}
\begin{equation}\label{D}
D_{J\,M_J}^{M_S}(t)=\frac{p_{-1}}{\sqrt{p_{-1}^2+p_1^2}}\sin\Bigl(\sqrt{2(p_{-1}^2+p_1^2)}\alpha
t\Bigr),
\end{equation}
\begin{equation}\label{E}
E_{J\,M_J}^{M_S}(t)=\frac{p_1}{\sqrt{p_{-1}^2+p_1^2}}\sin\Bigl(\sqrt{2(p_{-1}^2+p_1^2)}\alpha
t\Bigr).
\end{equation}
We are interested in the entanglement dynamics of the two central
spins $A$ and $B$. Tracing $\rho(t)=|\psi(t)\rangle\langle\psi(t)|$,
over all the degrees of freedom of the $M-2$ spins around $A$ and
$B$ we get the reduced density matrix of the central system. It
assumes the block diagonal structure given by eq.(\ref{struttura2}),
every matrix element being an explicit function of the probability
amplitudes (\ref{A})-(\ref{E}). In particular if $M_S=1$
\begin{equation}\label{ms1}
a(t)=|A_{J\,M_J}^{M_S}(t)|^2,\,b(t)=|C_{J\,M_J}^{M_S}(t)|^2,\,e(t)=|B_{J\,M_J}^{M_S}(t)|^2,
\end{equation}
while if $M_S=0$
\begin{equation}\begin{split}\label{ms0}
&a(t)=|D_{J\,M_J}^{M_S}(t)|^2,\,b(t)=|A_{J\,M_J}^{M_S}(t)|^2+|B_{J\,M_J}^{M_S}(t)|^2,\\
&e(t)=|E_{J\,M_J}^{M_S}(t)|^2
\end{split}\end{equation} and
\begin{equation}\label{ms-1}
a(t)=|B_{J\,M_J}^{M_S}(t)|^2,\,b(t)=|C_{J\,M_J}^{M_S}(t)|^2,\,e(t)=|A_{J\,M_J}^{M_S}(t)|^2
\end{equation}
if $M_S=-1$. Suppose to prepare the two spins of the central system
and the remaining $M-2\equiv N$ spins in the state
\begin{equation}\label{initial}
|\psi(0)\rangle=|1, 1\rangle |N/2, -N/2+k,1\rangle
\end{equation}
where $k=0,1,\cdots N$. Thanks to our results expressed by
eq.(\ref{cns1}), we can state with certainty that at a generic time
instant $t$ the two central spins are entangled if and only if
$\bigl(\Delta S_z\bigr)^2<2b(t)$. In Fig.(\ref{Immagine7a}) we plot
$(\Delta S_z)^2$ and $2b(t)$ in correspondence to $N=100$ and $k=2$
as a function of $\alpha t$. Since condition (\ref{cns1}) is not
verified whatever the time instant $t$ is, we can easily conclude
that starting from the initial condition (\ref{initial}) with
$N=100$ and $k=2$, the central system is unable to develop quantum
correlations. At the light of the result before discussed, the
physical reason of this incapacity of the system to develop
entanglement stems from the fact that the time evolution from this
initial condition never makes significantly different the
probabilities of finding the two spins in the state
$|\!\!\uparrow\uparrow\rangle$ or $|\!\!\downarrow\downarrow\rangle$
respectively.
\begin{figure}[hp]
\begin{center}
\includegraphics[scale=0.7]{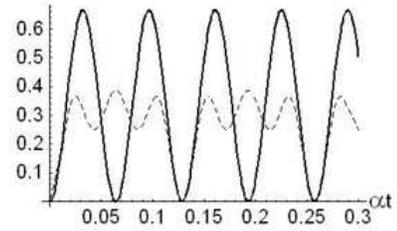}
\end{center}
\caption{$(\Delta S_z)^2$ (dashed line) and $2b(t)$ (bold line)
against $\alpha t$ in correspondence to $k=98$ and
$N=100$}\label{Immagine8a}
\end{figure}

\begin{figure}[h]
\begin{center}
\includegraphics[scale=0.55]{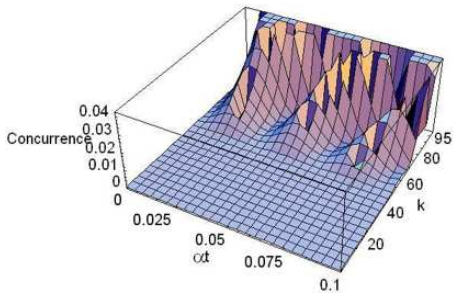}
\end{center}
\caption{Concurrence function of the central system against $\alpha
t$ and $k$ in correspondence to the initial condition $|1, 1\rangle
|N/2, -N/2+k\rangle$   for $N=100$}\label{Immagine1a}
\end{figure}

\begin{figure}[h]
\begin{center}
\includegraphics[scale=0.55]{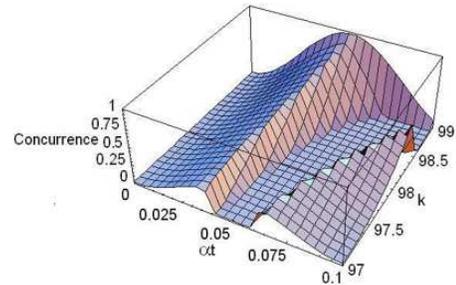}
\end{center}
\caption{Concurrence function of the central system against $\alpha
t$ and $k$  $({97\leq k <100})$ in correspondence to the initial
condition $|1, 1\rangle |N/2, -N/2+k\rangle$ for
$N=100$}\label{Immagine9a}
\end{figure}

It is of relevance the fact that the same conclusion is reached
starting from $|\psi(0)\rangle=|1, 1\rangle |N/2, -N/2+k,1\rangle$
with $k\leq50$. The behaviour of the system is instead deeply
different when $k$ exceeds $50$. As an example we compare in
Fig.(\ref{Immagine8a}) $(\Delta S_z)^2$ and $2b(t)$ for $k=98$ and
$N=100$. In this case, there exist different time intervals, in
which the two spins $A$ and $B$ are entangled because $(\Delta
S_z)^2<2b(t)$. Thus the parameter $k$ controlls the ability of the
system to generate entanglement in the central system. We expect
that the amount of entanglement also depends on the choice of $k$
that in turn determines the $S_z$ fluctuations. This indeed is true
as witnessed by Figures (\ref{Immagine1a}) and (\ref{Immagine9a})
where the time evolution of the concurrence is reported. We stress
that for $k=100$ the initial condition is an exact stationary state
while for $k=99$ the two central spins reach the maximum compatible
degree of entanglement since in this case the population of
$|\!\!\downarrow\downarrow\rangle$ exactly vanishes at any time
instant so that eq.(\ref{cns1}) is always fulfilled except when
$b(t)=0$. On this basis we foresee and we have proved a specular
behaviour starting from the initial condition $|1, -1\rangle |N/2,
-N/2+k,1\rangle$ in the sense that in this case the entanglement
reaches its maximum when $k=1$.
\vspace{2.5cm}
We emphasize that the constraint on the fluctuations of $S_z$
expressed by eq.(\ref{cns1}) is the key of the simplicity with which
we analyze the appearance and the disappearance of entanglement as a
function of time and more important to understand its dependence on
the value of $k$. In this paper we have proved that when two
two-level systems are described by a density matrix expressed by
eq.(\ref{struttura2}) at any time instant $t$, there is a simple and
reliable procedure of experimental interest after which the
occurrence of quantum correlations be surely claimed or excluded. We
propose indeed the measurement of at most two populations which
amounts at comparing the variance of $S_z$ with the probability of
finding one spin up and the other down. Our approach to the
entanglement of the pair of two level systems has the merit of
directly involving quantities having a clear physical meaning.
Applying our criterion to a spin star system we are able to fully
exploit the novelty of our point of view to explain the dependence
of the ability of the system to develop entanglement on the initial
conditions.

\end{document}